\documentclass[showpacs,preprintnumbers,superscriptaddress]{revtex4}
\usepackage{CJK}
\usepackage{amsmath,amssymb,graphicx,bm}
\begin{document}
\title{Accretion onto a Moving Reissner-Nordstr\"{o}m Black Hole}

\author{Lei Jiao}
\email{yangrongjia@tsinghua.org.cn}
\affiliation{College of Physical Science and Technology, Hebei University, Baoding 071002, China}
\author{Rongjia Yang\footnote{Corresponding author}}
\email{yangrongjia@tsinghua.org.cn}
\affiliation{College of Physical Science and Technology, Hebei University, Baoding 071002, China}
\affiliation{Hebei Key Lab of Optic-Electronic Information and Materials,Hebei University, Baoding 071002, China}

\begin{abstract}
We obtain an analytic solution for accretion of a gaseous medium with a adiabatic equation of state ($P=\rho$) onto a Reissner-Nordstr\"{o}m black hole which moves at a constant velocity through the medium. We obtain the specific expression for each component of the velocity and present the mass accretion rate which depends on the mass and the electric charge. The result we obtained may be helpful to understand the physical mechanism of accretion onto a moving black hole.
\end{abstract}

\pacs{97.60.Lf, 04.20.Jb, 47.75.+f, 95.30.Sf}

\maketitle

\section{Introduction}

Accretion of matter onto astronomical objects is a long-standing interesting phenomenon for astrophysicists. Pressure-free gas being dragged onto a star moving at constant velocity was first discussed qualitatively in \cite{Bondi:1944jm}. In the case of spherical accretion onto a stationary black hole, exact solutions have been found in the context of Newtonian gravity \cite{Bondi:1952ni} and in the framework of general relativity \cite{michel1972accretion}. Thereafter accretion has been analyzed in literatures for various black holes, such as a Schwarzschild \citep{Babichev:2004yx}, a charged black hole \citep{michel1972accretion,Jamil:2008bc}, a Kerr-Newman black hole \citep{JimenezMadrid:2005rk,Babichev:2008dy,Bhadra:2011me}, a Reissner-Nordstrom black hole \citep{Babichev:2008jb}, a Kiselev black hole \citep{Yang:2016sjy}, a higher dimensional black hole \citep{Giddings:2008gr,Sharif:2011ih,John:2013bqa,Debnath:2015yva}, a black hole in a string cloud background \citep{Ganguly:2014cqa}, and a cosmological black hole or a Schwarzschild-(anti-)de Sitter black hole \citep{Mach:2013fsa, Mach:2013gia, Karkowski:2012vt, Gao:2008jv}. An exact solution was derived for one or two dust shells collapsing towards a black hole \cite{Liu:2009ts}. Quantum gravity corrections to accretion onto a Schwarzschild black hole were investigated in \cite{Yang:2015sfa}. In general, numerical analysis are required to dealt with nonspherical accretion for either Newtonian or relativistic flow. One exact, fully relativistic, nonspherical solution had been found for a Schwarzschild or Kerr black hole moving through a medium obeying a stiff $P=\rho$ equation of state \citep{petrich1988accretion,1989ApJ}. This solution provides valuable insight into the more general cases and serves as a benchmark for the test of numerical codes. However, because of the mathematical complexity, until now analogous exact solutions still had not been obtained for a moving Reissner-Nordstr\"{o}m black hole. Here we consider accretion onto such a black hole, and find an exact, fully relativistic, nonspherical solution. The result obtained here may be helpful to understand the physical mechanism of accretion onto a moving black hole.

The rest of the paper is organized as follows. In next section, we will present the fundamental equations for accretion on to moving black hole. In section III and IV, we will determine the mass accretion rate for a moving Reissner-Nordstr\"{o}m black hole. Finally, we will briefly summarize and discuss our results in section V.

\section{Basic Equations}
We consider the flow in the black hole rest frame and seek a stationary solution, assuming a homogeneous fluid moving at constant velocity at large distances. The flow of matter is approximated as a perfect fluid. The relativistic vorticity tensor is defined as \citep{petrich1988accretion}
\begin{eqnarray}
\label{1}
\omega_{\mu\nu}=P^{\alpha}_{\mu}P^{\beta}_{\nu}\left[(hu_{\alpha})_{;\beta}-(hu_{\beta})_{;\alpha}\right],
\end{eqnarray}
where $;$ denotes the covariant derivative with respect to the coordinate, $u^{\mu}$ is the four-velocity, $P_{\mu}^{\nu}=\delta_{\mu}^{\nu}+u_{\mu}u^{\nu}$ is the projection tensor, and $h\equiv(\rho+P)/n$ is the enthalpy. We take the units $c=G=1$, $c$ and $G$ are the speed of light and the Newtonian gravitational constant, respectively. For the perfect fluid, Euler's equations reads
\begin{eqnarray}
\label{2}
(hu_{\mu})_{;\alpha}u^{\alpha}+h_{,\mu}=0.
\end{eqnarray}
Combining equations (\ref{1}) and (\ref{2}), we derive a simple expression for the vorticity:
\begin{eqnarray}
\label{3}
\omega_{\mu\nu}=[(hu_{\mu})_{;\nu}-(hu_{\nu})_{;\mu}]
\end{eqnarray}
As in Newtonian flow, the vorticity will be zero everywhere if it is zero on some initial hypersurface, and while if the vorticity is zero, the velocity of the perfect fluid can be expressed as the gradient of a potential \citep{1980ApJ}:
\begin{eqnarray}
\label{4}
hu_{\mu}=\psi_{,\mu}.
\end{eqnarray}
If no particles are created or destroyed, the equation of continuity for the particle density $n$ is
\begin{eqnarray}
\label{5}
(nu^{\alpha})_{;\alpha}=0,
\end{eqnarray}
or
\begin{eqnarray}
\label{6}
[(n/h)\psi^{,\alpha}]_{;\alpha}=0.
\end{eqnarray}
Following from Eq. (\ref{4}). the quantity $h$ is found from the normalization equation
\begin{eqnarray}
\label{7}
h=(-\psi^{,\alpha}\psi_{,\alpha})^{1/2}.
\end{eqnarray}
In general, equation (\ref{6}) is a nonlinear equation in $\psi$ and its derivatives. However, if $h\propto n$, equation (\ref{6}) becomes a linear equation. Like in \citep{petrich1988accretion}, we consider the simplest case $P=\rho\propto n^{2}$ which implies that the adiabatic index is equal to 2 and that the speed of sound is equal to the speed of light. The flow velocity must be subsonic everywhere and therefore no shock waves arise. Then we have to solve the equation
\begin{eqnarray}
\label{8}
(\psi^{,\alpha})_{;\alpha}=0,
\end{eqnarray}
with appropriate boundary conditions. An important result that we hope to obtain is the particle accretion rate
\begin{eqnarray}
\label{9}
\dot{N}=-\int_{S}nu^{i}\sqrt{-g}dS_{i}=-\int_{S}\psi_{,r}g^{rr}\sqrt{-g}d\Omega.
\end{eqnarray}
where $S$ is the boundary two-surface of a sphere centered on the black hole, $g^{rr}$ is radius-radius component of contravariant metric tensor, $g$ is the determinant of the metric.

Since the medium is homogeneous at large distances, we can set $n=h =1$ in appropriate units and restore $n_{\infty}$ later. In rectangular coordinates the asymptotic boundary condition is
\begin{eqnarray}
\label{10}
\psi=u_{\mu} x^{\mu}=-u_{\infty}^{0}t+\textbf{u}_{\infty}\cdot\textbf{x},
\end{eqnarray}
or in spherical coordinates
\begin{eqnarray}
\label{11}
\psi=-u_{\infty}^{0}t+u_{\infty}r[\cos\theta\cos\theta_{0}+\sin\theta\sin\theta_{0}\cos(\phi-\phi_{0})],
\end{eqnarray}
for $r\rightarrow\infty$. The asymptotic three-velocity vector $v_{\infty}$ can be allowed to point along an arbitrary direction $(\theta_{0},\phi_{0})$. Note that
\begin{eqnarray}
\label{12}
u_{\infty}^{\mu}=(u_{\infty}^{0},\textbf{u}_{\infty})=(1-v_{\infty}^{2})^{-1/2}(1,\textbf{v}_{\infty}).
\end{eqnarray}
Here we consider the flow is into, and not out from, a black hole. Other boundary conditions is that $n$ and $h$ be finite everywhere, including at the event horizon of a black hole.

\section{Accretion onto a Moving Reissner-Nordstr\"{o}m Black Hole}
We first consider accretion onto a static and spherically symmetric Reissner-Nordstr\"{o}m black hole moving at a constant velocity through the medium. The metric of the Reissner-Nordstr\"{o}m space-time is given by
\begin{eqnarray}
\label{13}
ds^{2}=-\left(1-\frac{2M}{r}+\frac{Q^{2}}{r^{2}}\right)dt^{2}+\frac{1}{1-\frac{2M}{r}+\frac{Q^{2}}{r^{2}}}dr^{2}+r^{2}d\theta^{2}+r^{2}\sin^{2}\theta d\phi^{2},
\end{eqnarray}
where $M$ is the mass of the black hole measured by an observer at infinity. $Q$ is the electric charge. If $e^2 > M^2$, the metric is non-singular everywhere except at the curvature singularity ($r = 0$). We consider the case of $|Q|<M $, the black hole has inner and outer horizon which are localized at $r_{\pm}=M\pm\sqrt{M^{2}-Q^{2}}$, respectively. The outer horizon denoted as $r_+$ is effectively called the event horizon.

For Reissner-Nordstr\"{o}m black hole, equation (\ref{8}) gives
\begin{eqnarray}
\label{14}
-\frac{1}{1-\frac{2M}{r}+\frac{Q^{2}}{r^{2}}}\frac{\partial^{2}\psi}{\partial t^{2}}+\frac{1}{r^{2}}\frac{\partial}{\partial r}\left[\left(1-\frac{2M}{r}+\frac{Q^{2}}{r^{2}} \right)r^{2}\frac{\partial\psi}{\partial r} \right]+\frac{1}{r^{2}}\left[\frac{1}{\sin\theta}\frac{\partial}{\partial\theta}\left(\sin\theta\frac{\partial\psi}{\partial\theta}\right)
+\frac{1}{\sin^{2}\theta}\frac{\partial^{2}\psi}{\partial\phi^{2}} \right]=0.
\end{eqnarray}
Considering the asymptotic boundary condition in spherical coordinates: $\psi=-u_{\infty}^{0}t+u_{\infty}r[\cos\theta\cos\theta_{0}+\sin\theta\sin\theta_{0}\cos(\phi-\phi_{0})]~(r\rightarrow\infty)$, we can assume the general formula of the solution takes the form: $\psi=-u_{\infty}^{0}t+u(r,\theta,\phi)$, it should satisfy the stationary flow condition that the gradient of $\psi$ should be independent of time. Substituting $\psi$ in the equation (\ref{14}) with the general form, we obtain
\begin{eqnarray}
\label{15}
\frac{\partial}{\partial r}\left[\left(1-\frac{2M}{r}+\frac{Q^{2}}{r^{2}}\right)r^{2}\frac{\partial}{\partial r}u\right]+\frac{1}{\sin\theta}\frac{\partial}{\partial\theta}\left(\sin\theta\frac{\partial}{\partial\theta}u\right)
+\frac{1}{\sin^{2}}\frac{\partial^{2}}{\partial\phi^{2}}u=0.
\end{eqnarray}
This is the differential equation that the spacial part $u(r,\theta,\phi)$ of $\psi$ satisfies. Assume $u=R(r)\Theta(\theta)\Phi(\phi)$, the functions $R(r)$, $\Theta(\theta)$, $\Phi(\phi)$ satisfy respectively the following equations
\begin{eqnarray}
\label{16}
\frac{d^{2}}{d\phi^{2}}\Phi+m^{2}\Phi=0,
\end{eqnarray}
\begin{eqnarray}
\label{17}
\frac{1}{\sin\theta}\frac{d}{d\theta}\left(\sin\theta\frac{d}{d\theta}\Theta \right)-\frac{m^{2}}{\sin^{2}\theta}\Theta+l(l+1)\Theta=0,
\end{eqnarray}
and
\begin{eqnarray}
\label{18}
\frac{d}{dr}\left[\left(1-\frac{2M}{r}+\frac{Q^{2}}{r^{2}}\right)r^{2}\frac{d}{dr}R\right]-l(l+1)R=0,
\end{eqnarray}
Let $r=M+\xi\sqrt{M^{2}-Q^{2}}$, then Eq. (\ref{18}) can be rewritten as
\begin{eqnarray}
\label{19}
(1-\xi^{2})\frac{d^{2}R}{d\xi^{2}}-2\xi\frac{dR}{d\xi}+l(l+1)R=0,
\end{eqnarray}
This is the Legendre equation. The general solutions for $\Phi(\phi)$, $\Theta(\theta)$, $R(\xi)$ can take the following forms, respectively
\begin{eqnarray}
\label{20}
\Phi(\phi)=C\cos m\phi+D\sin m\phi,~~m=0,1,2,...,
\end{eqnarray}
\begin{eqnarray}
\label{21}
\Theta(\theta)=P_{l}^{m}(cos\theta),~~m,l=0,1,2,...,
\end{eqnarray}
\begin{eqnarray}
\label{22}
R(\xi)=AP_{l}(\xi)+BQ_{l}(\xi),~~l=0,1,2,...,
\end{eqnarray}
where $A$, $B$, $C$, and $D$ are constants, $P_{l}^{m}(z)$ is the associated Legendre function, $P_{l}(z)$ is Legendre polynome, and $Q_{l}(z)$ is the second kind of Legendre function which is linearly independent from $P_{l}(z)$. Therefore the general formula of $\psi$ for Reissner-Nordstr\"{o}m black hole is
\begin{eqnarray}
\label{23}
\psi=-u^{0}_{\infty}t+\sum_{l,m}\left[A_{lm}P_{l}(\xi)+B_{lm}Q_{l}(\xi)\right]Y_{lm}(\theta,\phi),
\end{eqnarray}
where $A_{lm},B_{lm}$ are constants to be determined from the boundary conditions, $Y_{lm}(\theta,\phi)$ are spheric harmoics, which are made up of $\Phi(\phi)$ and $\Theta(\theta)$. From equation (\ref{4}), the velocities are given by
\begin{eqnarray}
\label{24}
nu_{t}=-u^{0}_{\infty},
\end{eqnarray}
\begin{eqnarray}
\label{25}
nu_{r}=\frac{1}{\sqrt{M^{2}-Q^{2}}}\sum_{l,m}[A_{lm}P_{l}^{'}(\xi)+B_{lm}Q_{l}^{'}(\xi)]Y_{lm}(\theta,\phi),
\end{eqnarray}
\begin{eqnarray}
\label{26}
nu_{\theta}=\sum_{l,m}[A_{lm}P_{l}(\xi)+B_{lm}Q_{l}(\xi)]\frac{\partial Y_{lm}(\theta,\phi)}{\partial\theta},
\end{eqnarray}
\begin{eqnarray}
\label{27}
nu_{\phi}=\sum_{l,m}[A_{lm}P_{l}(\xi)+B_{lm}Q_{l}(\xi)]\frac{\partial Y_{lm}(\theta,\phi)}{\partial\phi},
\end{eqnarray}
where the prime $'$ denotes the derivative with respect to $\xi$. The normalization condition yieldes
\begin{eqnarray}
\label{28}
n^{2}=\frac{1}{1-\frac{2M}{r}+\frac{Q^{2}}{r^{2}}}(u^{0}_{\infty})^{2}-\left(1-\frac{2M}{r}+\frac{Q^{2}}{r^{2}}\right)(nu_{r})^{2}
-\frac{1}{r^{2}}(nu_{\theta})^{2}-\frac{1}{r^{2}\sin^{2}\theta}(nu_{\phi})^{2}.
\end{eqnarray}
An important constraint is the finiteness of $n$ at the outer horizon, $r_{+}=M+\sqrt{M^{2}-Q^{2}}$, where $\xi=1$. Using the limiting behavior of the Legendre functions near the outer horizon, we have
\begin{eqnarray}
\label{29}
n^{2}\rightarrow\frac{1}{1-\frac{2M}{r}+\frac{Q^{2}}{r^{2}}}\left\{(u^{0}_{\infty})^{2}-\left[\frac{\sqrt{M^{2}-Q^{2}}}{(M+\sqrt{M^{2}-Q^{2}})^{2}}\sum_{l,m}B_{lm}Y_{lm}(\theta,\phi)\right]^{2}\right\},
\end{eqnarray}
which implies that
\begin{eqnarray}
\label{30}
B_{00}Y_{00}=u^{0}_{\infty}\frac{(M+\sqrt{M^{2}-Q^{2}})^{2}}{\sqrt{M^{2}-Q^{2}}},
\end{eqnarray}
with all the other $B$'s vanish. For inward accretion, we select the positive sign for $B_{00}$. Equation (\ref{23}) reduces to
\begin{eqnarray}
\label{31}
\psi=-u^{0}_{\infty}t-u^{0}_{\infty}\frac{(M+\sqrt{M^{2}-Q^{2}})^{2}}{2\sqrt{M^{2}-Q^{2}}}\ln{\frac{r-r_{+}}{r-r_{-}}}+\sum_{l,m}A_{lm}P_{l}(\xi)Y_{lm}(\theta,\phi),
\end{eqnarray}
where the $A$ can now be found from the asymptotic boundary conditions in Eq.(\ref{11}).  Without loss of generality, we consider that the flow is accreted toward the north pole of the coordinate system: $\theta=0$. Then the asymptotic boundary condition turns to
\begin{eqnarray}
\label{32}
\psi=-u^{0}_{\infty}t+u_{\infty}r\cos\theta.
\end{eqnarray}
Comparing it with Eq. (\ref{31}), we have
\begin{eqnarray}
\label{33}
A_{10}=u_{\infty}\sqrt{M^{2}-Q^{2}},
\end{eqnarray}
with all the other $A_{lm}$'s zero. Then we obtain the final solution
\begin{eqnarray}
\label{34}
\psi=-u^{0}_{\infty}t-u^{0}_{\infty}\frac{(M+\sqrt{M^{2}-Q^{2}})^{2}}{2\sqrt{M^{2}-Q^{2}}}\ln{\frac{r-r_{+}}{r-r_{-}}}+u_{\infty}(r-M)\cos\theta.
\end{eqnarray}
Each component of the velocity is given by
\begin{eqnarray}
\label{35}
nu_{t}=-u^{0}_{\infty},
\end{eqnarray}
\begin{eqnarray}
\label{36}
nu_{r}=-u^{0}_{\infty}\frac{(M+\sqrt{M^{2}-Q^{2}})^{2}}{r^{2}-2Mr+Q^{2}}+u_{\infty}\cos\theta,
\end{eqnarray}
\begin{eqnarray}
\label{37}
nu_{\theta}=-u_{\infty}(r-M)\sin\theta,
\end{eqnarray}
\begin{eqnarray}
\label{38}
nu_{\phi}=0.
\end{eqnarray}
The density takes the form
\begin{eqnarray}
\label{39}
\nonumber
n^{2}&=&(u^{0}_{\infty})^{2}\frac{1}{1-\frac{M-\sqrt{M^{2}-Q^{2}}}{r}}\left[1+\frac{M+\sqrt{M^{2}-Q^{2}}}{r}+\left(\frac{M+\sqrt{M^{2}-Q^{2}}}{r}\right)^{2}
+\left(\frac{M+\sqrt{M^{2}-Q^{2}}}{r}\right)^{3}\right]\\
&-&(u_{\infty})^{2}\left[\left(1-\frac{2M}{r}\right)+\left(\frac{M}{r}\right)^{2}\sin^{2}\theta+\frac{Q^{2}}{r^{2}}\cos^{2}\theta\right]
+\frac{2(M+\sqrt{M^{2}-Q^{2}})^{2}}{r^{2}}u^{0}_{\infty}u_{\infty}\cos\theta.
\end{eqnarray}
We can judge from this formula that $n$ is finite at the outer horizon. The point at which the velocity is zero is called a stagnation point, lies at $\theta=0$ (directly downstream) and at radius
\begin{eqnarray}
\label{40}
r=M+\left[M^{2}-Q^{2}+\frac{1}{v_{\infty}}(M+\sqrt{M^{2}-Q^{2}})^{2}\right]^{1/2}.
\end{eqnarray}
From Eq. (\ref{9}), we derive the accretion rate (restoring $n_{\infty}$)
\begin{eqnarray}
\label{41}
\dot{N}=4\pi u^{0}_{\infty}n_{\infty}r^{2}_+.
\end{eqnarray}
This is just the area of the outer horizon of the black hole multiplies $n_{\infty}$ and Lorentz factor $\gamma$ for the flow at large distances. If $Q=0$, equation (\ref{41}) reduces to the accretion rate for Schwarzschild black hole \citep{petrich1988accretion,1989ApJ}.

\section{Conclusions and discussions}
We have considered accretion onto a Reissner-Nordstr\"{o}m black hole. The black hole moves at a constant velocity through the medium which obeys a stiff $P=\rho$ equation of state. We obtained the mass accretion rate which depends on the mass and the electric charge. We obtain the specific expression for each component of the velocity which implies the flow is two spatial dimension. The results obtained here may provide valuable physical insight into the more complicated cases and can be generalized to other types black hole.

\begin{acknowledgments}
We thank Stuart L. Shapiro for helpful discussions. This study is supported in part by National Natural Science Foundation of China (Grant Nos. 11147028 and 11273010), the Hebei Provincial Outstanding Youth Fund (Grant No. A2014201068), the Outstanding Youth Fund of Hebei University (No. 2012JQ02), and the Midwest universities comprehensive strength promotion project.
\end{acknowledgments}

\bibliography{ref}

\end{document}